\documentclass[a4paper,11pt]{article}
\usepackage{pos}
\usepackage{graphicx}
\usepackage{slashed}
\usepackage{bm}

\title{Renormalon-chain contributions to two-point correlators of nonlocal quark currents and light-meson distribution amplitudes}
 \ShortTitle{Renormalon chains and distribution amplitudes}

\author[a]{S.~V.~Mikhailov}
\author*[a]{N.~Volchanskiy}

\affiliation[a]{Bogoliubov Laboratory of Theoretical Physics, Joint Institute for Nuclear Research,\\ 141980 Dubna, Russia
}

\emailAdd{mikhs@theor.jinr.ru}
\emailAdd{nikolay.volchanskiy@gmail.com}

\abstract{
In the framework of QCD sum rules, we study the endpoint slopes of the leading-twist distribution amplitudes in the large-$\beta_0$ approximation for the pion and the longitudinally polarized $\rho$ meson. To this end, we compute the two-point massless correlator of non-local composite vector quark currents with the gluon lines dressed by chains of one-loop fermion bubbles. This correlator depends on the Bjorken fraction $x$ associated with the composite current and, within the large-$\beta_0$ approximation, provides the dominant contribution in each order of perturbation theory. We derive an inequality for the second Gegenbauer moments of these distribution amplitudes that agrees  with predictions from lattice QCD.
}

\FullConference{43rd International Conference on High Energy Physics (ICHEP 2026)\\
30 July  to 5 August, 2026\\
Natal, Brazil\\}

\begin{document}
\maketitle

\section{Introduction}

In the factorization approach to describing hard exclusive hadronic processes in QCD, meson distribution amplitudes (DAs) encapsulate the soft (long-distance) parton dynamics. When combined in integral convolutions with purely perturbative (short-distance) coefficient functions, they govern the behavior of the corresponding form factors (FFs) and amplitudes \cite{Efremov:1979qk}. DAs can be computed using a variety of theoretical tools, including lattice QCD (LQCD) and QCD sum rules (SRs)---the latter being the focus of this paper. For the lightest pseudoscalar and vector mesons---the pion and the longitudinally polarized $\rho$ meson---the leading-twist-2 QCD SRs for their DAs, $\varphi_\text{mes}(x)$, read \cite{Mikhailov:2021kty}
\begin{align}\label{eq:sr}
f_\text{mes}^2  e^{-m_\text{mes}^2/M^2}\varphi_\text{mes}(x) + \ldots
=
\Phi_\text{PT}(x;M;s_0^\text{mes}) \pm \Phi_\text{S} (x;M) + \Delta_\text{nSC}(x,M) \stackrel{\text{def}}{=} \Phi_\text{mes}(x,M).
\end{align}
Here, the upper sign corresponds to $\text{mes} = \pi$ in the chiral limit $m_\pi = 0$ and the lower sign to $\text{mes} = \rho^\|$; $M^2$ denotes the Borel parameter, and $x\in[0,1]$ is the momentum fraction (Bjorken variable) carried by the valence quark. On the left-hand side (l.h.s.)\ of the SR \eqref{eq:sr}, the ellipses indicate the exponentially suppressed contributions from heavier meson states. The right-hand side (r.h.s.)\ of the SR, $\Phi_\text{mes}(x,M)$, comprises two components. The first component is dominant and arises from the perturbative QCD (pQCD) spectral density $\rho_\text{pt}(x;s)$:
\begin{align} \label{eq:PT-rho}
\Phi_\text{PT}(x;M;s_0 ) = \int_{0}^{s_0} \rho_\text{pt}(x;s)e^{-s/M^2}ds,
\end{align}
where the continuum threshold $s_0$ determines the duality intervals for the axial ($\pi$) and vector ($\rho^\|$) channels, respectively. The second component, $\pm \Phi_\text{S} + \Delta_\text{nSC}$, stems from the operator product expansion (OPE) formulated via non-local condensates (NLC). 

\begin{figure}[ht]
\begin{gather*}
\rho_\text{pt}(x;s) = \frac{N_c}{2\pi} x\bar{x} + i\sum_{n \geqslant 0}
\mathbf{\hat{B}^2} \, \begin{gathered}\includegraphics[width=.12\textwidth,bb= 0in 0in .642in 0.49in]{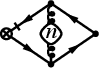}\end{gathered}
+\ldots
\qquad\qquad
\Phi_\text{S} \sim i\sum_{n \geqslant 0} \mathbf{\hat{B}} \, \begin{gathered}\includegraphics[width=.12\textwidth,bb= 0in 0in .642in 0.49in]{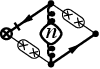}\end{gathered}
+ \ldots 
\end{gather*}
\caption{\label{fig:renorm+cond} \footnotesize Left: The renormalon-chain contribution to the perturbative spectral density $\rho_\text{pt}$ in Eq.~\eqref{eq:PT-rho}. Right: The scalar NLC piece $\Phi_\text{S}(x;M)$. The two ovals depict the scalar NLC evaluated within the vacuum factorization approximation for four-quark NLCs. The hard propagators of the coefficient function are highlighted with thicker lines. The ellipses stand for other diagrams of the same order. The Borel transform $\mathbf{\hat{B}}$ and its second iterate $\mathbf{\hat{B}}^2$ are defined in Ref.~\cite{Mikhailov:2023lqe}.}
\end{figure}

The structure of the perturbative coefficient functions for light-meson FFs implies that the most quantitatively important regions of the DA lie in the vicinity of the endpoints $x=0$ and $x=1$. Near these boundaries, the scalar NLC $\Phi_\text{S}$ dominates over all other non-leading NLC corrections contained in $\Delta_\text{nSC}$ \cite{Bakulev:2001pa}. Remarkably, the sign of this dominant nonperturbative term is the sole distinction between the OPE sectors of the QCD SRs in the vector and axial-vector channels \cite{Mikhailov:2021kty}. Consequently, to analyze the endpoint behavior of these DAs and to derive a direct relation between them, we retain only these two dominant components: $\Phi_\text{PT}$ and $\Phi_\text{S}$. The underlying Feynman diagrams are illustrated schematically in Fig.~\ref{fig:renorm+cond}. Both terms are evaluated within the large-$\beta_0$ (renormalon) approximation, which corresponds to evaluating one-loop fermion bubble-chain corrections in gluon propagators. This abridged conference proceedings paper is based on Refs.~\cite{Mikhailov:2021kty, Mikhailov:2023lqe, Mikhailov:2018udp}, to which we refer the reader for technical derivations and results not explicitly reproduced here.

%=========================================================
%=========================================================
%=========================================================

\section{Perturbative term $\Phi_\text{PT}$ in the large-$\beta_0$ approximation}

The spectral density of the perturbative part $\Phi_\text{PT}$ on the r.h.s.\ of Eq.~(\ref{eq:sr}) can be obtained as the iterated Borel transform $\mathbf{\hat{B}}$ of an $x$-dependent correlator (see Ref.~\cite{Mikhailov:2023lqe} for the definitions of $\mathbf{\hat{B}}$ and~$\mathbf{\hat{B}}^2$):
\begin{align}\label{eq:cor-def}
\rho_\text{pt}(x;s)= \mathbf{\hat{B}}^2_{(s \to P^2)}\Pi(x,\underline{0};L),
\qquad
\Pi(x,\underline{0};L)
= i \int d^D \eta \, e^{ip\eta} \langle 0|\hat{\mathrm{T}}\left[J^\dagger(\eta;x)J(0;\underline{0}) \right]|0 \rangle,
\end{align}
where $\eta$ is a spacetime coordinate, $p$ is the external momentum, and $L=\ln(P^2/\mu^2)$ with $P^2=-p^2$. The composite vector quark current $J(\eta;x)$ is defined as the inverse Mellin transform of a quark bilinear containing the $N$-th covariant derivative of the quark field operator:
\begin{align}\label{eq:currents-def1}
J(\eta;\underline{N}) =  \int_0^1 x^N J(\eta;x) \, dx = \bar{q}(\eta) \slashed{\tilde{n}} \left(i\tilde{n}\mathscr{D}\right)^N q(\eta), 
\qquad 
i\mathscr{D}_\mu=i\partial_\mu + g t_a A^a_\mu,
\end{align}
where $\tilde{n}^\mu$ is a light-like vector satisfying $\tilde{n}^2=0$. In this paper, the arguments of the Mellin transform are underlined; therefore, $J(\eta;\underline{0})$ corresponds to the standard local vector quark current.

The correlator can be written as an integral over the correlator of two composite currents, $\Pi(x,y;L)$:
\begin{gather}\label{eq:pixy}
\Pi(x,\underline{0};L)=\int_0^1 \Pi(x,y;L) \, dy.
\end{gather}
In the large-$\beta_0$ approximation, the correlator $\Pi(x,y;L)$ can be computed using the two-loop ``kite'' master integral considered in Ref.~\cite{Mikhailov:2018udp}. 

The general expression for the $n$-bubble renormalon corrections to the correlator $\Pi(x,\underline{0};L)$ was derived in Ref.~\cite{Mikhailov:2023lqe} in terms of two generating functions: one for the correlator with a renormalized $n$-bubble chain, and another for the renormalization of the composite current. These generating functions contain no functions more complex than the hypergeometric function $_3F_2(x)$ or $_3F_2(\bar{x})$, where $\bar{x}=1-x$. Using these generating functions, one can derive explicit coefficients for the $L$-expansion of the correlator with an $n$-bubble chain, $\Pi_{n}(x,\underline{0};L)$\footnote{In QCD with the gauge group $\text{SU}(N_c)$ for $N_c=3$, the Casimir invariants are $C_A=N_c$ and $C_F=T_F(N_c^2-1)/N_c$, where $T_F = 1/2$. We also define $a_s(\mu^2) = \alpha_s(\mu^2)/(4\pi)$, $\alpha_s=g^2/(4\pi)$, and $\mu$ is the renormalization scale. The one-loop QCD $\beta$-function coefficient is $\beta_0 = \frac{11}{3} C_A - \frac43 T_F n_f = 9$ for $n_f=3$ massless quark flavors.}:
\begin{gather}\label{eq:PiSeries}
\mathbf{\hat{B}}\Pi(x,\underline{0};L) = \frac{N_c}{\pi^2} \left[ \frac12 x \bar{x}+ a_s C_F\sum_{n \geqslant 0} A^n \mathbf{\hat{B}}\Pi_{n}(x,\underline{0};L)\right],
\\
\Pi_n(x,\underline{0};L) = (-1)^n n! \sum_{k=0}^{n+1} \frac{(-L)^k}{k!} \Pi_n^k(x,\underline{0}),
\end{gather}
where the constant is $A = \frac43 a_s(\mu^2) T_F n_f$ or $-a_s \beta_0$ (corresponding to naive non-abelianization). The first non-vanishing coefficient reads
\begin{gather} \label{eq:leadingP_n}
\Pi_{n}^{n+1}(x,\underline{0}) = \frac12 \mathop{\hat{\mathbf{S}}} \Biggl\{ x \ln x
- \delta_{0,n} \left[x \ln x - \frac12 x \bar{x} \left( \frac{\pi^2}{3} - 5 - \ln^2 \frac{x}{\bar{x}} \right) \right] \Biggr\},
\;\;
 \mathop{\hat{\mathbf{S}}} f(x) = f(x) + f(\bar{x}),
\end{gather}
which agrees with the previous calculations for $n=0, 1$ in Ref.~\cite{Mikhailov:2018udp}. The subleading coefficients can be expressed in terms of the harmonic polylogarithms (HPLs) of Ref.~\cite{Remiddi:1999ew}, $\mathrm{H}_{\bm{\mu}}(x)$ \cite{Mikhailov:2023lqe}:
\begin{align}
\Pi_{n}^{k}(x,\underline{0}) \propto \mathop{\hat{\mathbf{S}}} \mathrm{H}_{\bm{\mu}}(x) + \text{lower-weight HPLs}
\qquad
\text{for } k \leqslant n, 
\end{align}
where the multi-index is $\bm{\mu}=\mu_1,\dots, \mu_r$ with $\mu_i>0$ and $\sum \mu_i = n-k+3$.

The series \eqref{eq:PiSeries} is asymptotic and is characterized by an increasingly singular behavior of $\hat{\mathbf{B}}\Pi_{n}(x,\underline{0};L)$ at the endpoints as $n$ grows \cite{Mikhailov:2023lqe}. By considering the zeroth and inverse moments, $\hat{\mathbf{B}}\Pi_{n}(\underline{0},\underline{0};L)$ and $\hat{\mathbf{B}}\Pi_{n}(\underline{-1},\underline{0};L)$ (which quantitatively capture the intermediate-$x$ and near-endpoint behavior, respectively), we find that the series \eqref{eq:PiSeries} should be truncated at $n=3$ or $4$, beyond which it exhibits factorial growth \cite{Mikhailov:2023lqe}.

The spectral density $\rho_\text{pt}$ in Eq.~\eqref{eq:cor-def} within the large-$\beta_0$ approximation is illustrated diagrammatically in the left panel of Fig.~\ref{fig:renorm+cond}. The first of these radiative corrections has been known for some time \cite{Mikhailov:1988nz}; see Eq.~(\ref{eq:leadingP_n}) for $n=0$.

%=================================================================================
%=================================================================================
%=================================================================================

\section{Pion DA with $a_s(a_s\beta_0)^3$ corrections}

\begin{figure}[ht]
\begin{center}
\includegraphics[width=.24\textwidth]{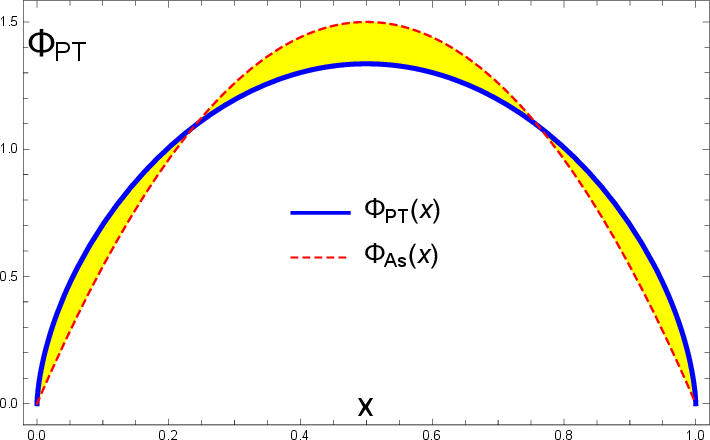}
\includegraphics[width=.24\textwidth]{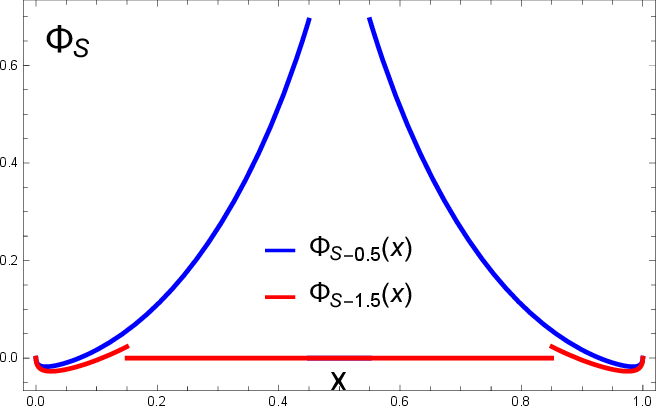}
\includegraphics[width=.24\textwidth]{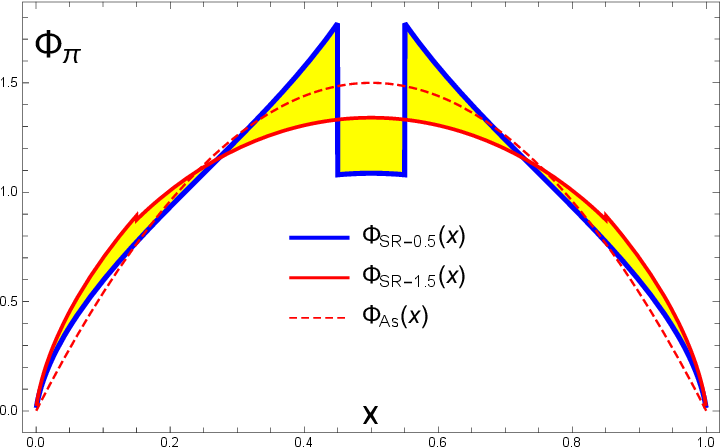}
\includegraphics[width=.24\textwidth]{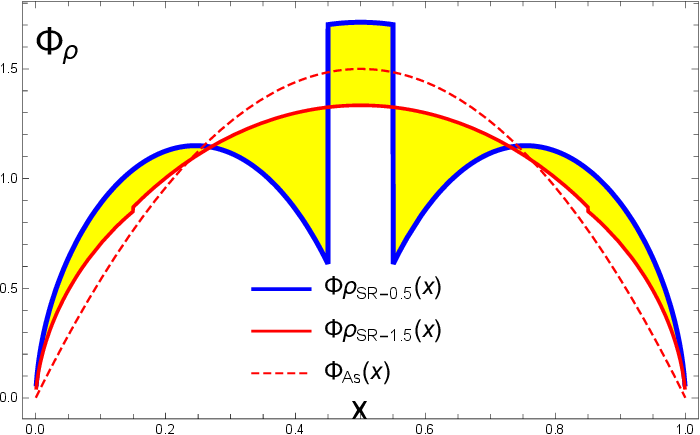}
\caption{\label{fig:qcdsr} \footnotesize 
Left panel: the perturbative part $\Phi_\text{PT}(x)$ of the r.h.s.\ of the NLC SR \eqref{eq:sr} (solid blue line) compared to the asymptotic profile $\Phi_\text{As}(x)$ (dashed red line). Second panel: the scalar NLC contribution $\Phi_\text{S}(x)$. Third panel: the total r.h.s.\ of the NLC SR for the pion, $\Phi_\pi$. Right panel: the total r.h.s.\ of the NLC SR for the longitudinally polarized $\rho$ meson, $\Phi_\rho(x)$. All contributions to the NLC QCD SRs are evaluated up to order $a_s(a_s\beta_0)^3$. The Borel parameter $M^2$ varies within the interval $[M^2_- = 0.5\text{ GeV}^2\text{ (blue lines)}, M^2_+ = 1.5\text{ GeV}^2\text{ (red lines)}]$.}
\end{center}
\end{figure}

The inclusion of higher-order $\beta_0 a_s$ corrections leads to a steeper rise of the component $\Phi_\text{PT}(x)$ near the endpoints. This broadening of the perturbative part of the DAs is clearly visible in the leftmost panel of Fig.~\ref{fig:qcdsr} in comparison with the asymptotic DA $\Phi_\text{As}(x) = 6 x \bar{x}$. As previously noted, the large-$\beta_0$ part of the perturbative series behaves well up to $\text{N}^5\text{LO}$ \cite{Mikhailov:2023lqe}. Here, we incorporate corrections up to order $a_s(a_s\beta_0)^3$, as the subsequent order yields a negligible numerical impact. The renormalon corrections to the NLC part $\Phi_\text{S}$, however, largely compete with the perturbative broadening, effectively canceling it out, as illustrated in the second and third panels of Fig.~\ref{fig:qcdsr}.

We employ the Gaussian model for the NLC \cite{Mikhailov:1988nz}. This model introduces a single vacuum scale---the average virtuality of vacuum quarks, $\langle k_{q}^{2} \rangle = \lambda_{q}^{2} \equiv \langle\bar{q}D^{2}q\rangle /\langle\bar{q}q\rangle\big|_{\mu_0^2\simeq 1\text{ GeV}^2} \approx 0.45\text{ GeV}^2$ \cite{Mikhailov:2021znq}. The Gaussian model omits the details of the vacuum quark-gluon distributions at the expense of introducing finite discontinuities in the nonperturbative part of the SR's r.h.s.\ (see Fig.~\ref{fig:qcdsr}). It should be emphasized, however, that the SR~\eqref{eq:sr} is to be understood in a weak sense---namely, for integral convolutions in $x$ on both sides of the QCD SR, with the Borel parameters restricted to the stability domain. Examples of such convolutions include the moments $\langle  \xi^N \rangle_\text{mes}$ (where $\xi=2x-1$) and the inverse moment $\langle x^{-1} \rangle_\text{mes}$. 

Diagrammatically, the calculation of $\Phi_\text{S}(x;M)$ is illustrated in Fig.~\ref{fig:renorm+cond}. The term $\Phi_\text{S}$ can be expressed as an integral over a pair of scalar NLCs contracted with a coefficient function \cite{Mikhailov:1988nz,Mikhailov:2010ud}, which now incorporates the renormalon chain. The Gaussian decay of the scalar NLC ensures that this integral converges and does not require additional renormalization.

%=====================================================================================
%=====================================================================================
%=====================================================================================

\section{Longitudinally polarized $\rho$-meson DA with $a_s(a_s\beta_0)^3$ corrections}

For the longitudinally polarized $\rho$ meson, the NLC contribution $\Phi_\text{S}$ enters the r.h.s.\ of the QCD SR \eqref{eq:sr} with a sign \textit{opposite} to that of the pion case, which amplifies the impact of the perturbative corrections near the endpoints. This causes the profile of $\varphi_{\rho}^{\|}(x)$ to broaden near the boundaries; this ``swelling'' of the amplitude profile is illustrated in the right panel of Fig.~\ref{fig:qcdsr}. The endpoint slope of $\varphi_{\rho}^{\|}(x)$ ranges between 9 and 12, which is significantly larger than the value of $6$--$7$ obtained for $\varphi_{\pi}(x)$. Consequently, the inverse moment is enhanced to $\langle x^{-1}\rangle_\rho \approx 3.8$, compared to $\langle x^{-1}\rangle_\pi \approx 3.4$.

Although obtaining reliable estimates for the higher Mellin moments $\langle \xi^n\rangle_\text{mes}$ is not feasible, we can still establish inequalities between the $\xi$-moments of the pion and $\rho$-meson DAs. Because the pion and $\rho^\|$ channels differ primarily in the sign of the dominant nonperturbative contribution---which originates from the scalar NLC---a direct relation between the two amplitudes can be derived by taking the difference of their respective sum rules. This approach yields inequalities that reflect the pronounced difference in the DA ``swelling'' described above:
\begin{align} 
\langle \xi^2\rangle_{\rho^{\|}}> \langle \xi^2\rangle_{\pi}> \langle \xi^2\rangle_\text{As}=\frac{1}{5}\,,
\qquad\text{or, equivalently,}\qquad
a_2^{\rho^{\|}} > a_2^{\pi}>0\,, \label{eq:ineq-a2}
\end{align}
where $a_2^\text{mes}=\frac{7}{12}\left(5\langle \xi^2\rangle_\text{mes}-1 \right)$ denotes the second Gegenbauer moment of the meson DA.

Lattice QCD calculations \cite{Braun:2016wnx} are in agreement with the inequalities \eqref{eq:ineq-a2}, yielding $a^{\rho^{\|}}_2(\mu_\text{L})=0.132(27) > a^{\pi}_2(\mu_\text{L})=0.116(20)$ at the scale $\mu_\text{L}^2=4\text{ GeV}^2$. Furthermore, the lattice result for the pion, $a^{\pi}_2(\text{Latt})$, is consistent with the QCD SRs prediction from Ref.~\cite{Mikhailov:2021znq}.
 
%=====================================================================================
%=====================================================================================
%=====================================================================================

\section{Conclusion}

In the large-$\beta_0$ approximation, we have computed the correlator $\Pi(x,\underline{0};L)$ of two vector quark currents, where one of the currents is composite. Our general result aligns with all known special cases available in the literature. We derived generating functions for this correlator and demonstrated that at any fixed order $a_s(a_s\beta_0)^n$, the correlator can be expressed in terms of harmonic polylogarithms of weight no higher than $n+3$. This allowed us to explore the asymptotic properties of the borelized correlator, showing that the perturbative series can be truncated at $n=3$ or $4$. 

Using these results, we estimated the large-$\beta_0$ contributions to QCD SRs for the DAs of the pion and the longitudinally polarized $\rho$ meson near the endpoints $x=0$ and $1$. While these contributions do not significantly alter the slope of the pion DA profile and remain fully consistent with experimental data on the pion transition form factor \cite{Mikhailov:2021znq}, they significantly modify the endpoint behavior of the $\rho$-meson DA. This modification reconciles the QCD SRs predictions with LQCD results regarding the hierarchy of the second Gegenbauer moments, $a_2^{\rho^{\|}} > a_2^{\pi}>0$ \cite{Braun:2016wnx}.

\section*{Acknowledgments}

N.V.\ is grateful to the management and colleagues at JINR and BLTP for their support and organizational efforts in facilitating travel to and participation in the conference. Special thanks also go to the organizers of ICHEP 2026 for the opportunity to present this talk and for hosting an excellent conference.

%====================================================================================

%\bibliographystyle{JHEP}
%\bibliography{conf-mod,refs,multiloop_calc}
%\begin{thebibliography}{99}
% \end{thebibliography}
\providecommand{\href}[2]{#2}\begingroup\raggedright
\endgroup
%[10]
 \end{document}